\documentclass[showpacs,preprintnumbers,pra,superscriptaddress,onecolumn]{revtex4}
\usepackage{amsmath}
\usepackage{graphicx}
\usepackage{dcolumn}
\usepackage{bm}

\begin{document}

\title{Diffusive teleportation on a quantum dot chain}
\author{F. de Pasquale}
\affiliation{INFM Center for Statistical Mechanics and Complexity}
\affiliation{Dipartimento di Fisica, Universit\`{a} di Roma La
Sapienza, P. A. Moro 2, 00185 Rome, Italy}

\author{G. Giorgi}\email{gianluca.giorgi@roma1.infn.it}
\affiliation{INFM Center for Statistical Mechanics and Complexity}
\affiliation{Dipartimento di Fisica, Universit\`{a} di Roma La
Sapienza, P. A. Moro 2, 00185 Rome, Italy}
\author{S. Paganelli}
\affiliation{Dipartimento di Fisica, Universit\`{a} di Roma La
Sapienza, P. A. Moro 2, 00185 Rome, Italy} \affiliation{Dipartimento
di Fisica, Universit\`{a} di Bologna, Via Irnerio 46, I-40126,
Bologna, Italy}

\pacs{03.67.Hk, 03.67.Mn, 85.35.Be}

\begin{abstract}
We introduce a model of quantum teleportation on a channel built on
a quantum dot chain. Quantum dots are coupled through hopping and
each dot can accept zero, one or two electrons. Vacuum and double
occupation states have the same potential energy, while single
occupation states are characterized by a lower potential energy. A
single dot initially decoupled from the others is weakly coupled
with an external element (Bob), where a pair of electrons has been
previously localized. Because of hopping after a suitable time the
two dots charge states become maximally entangled. Another chain dot
(Alice) is put in an unknown superposition of vacuum and double
occupation states, and the other dots are initially empty. The time
evolution of the system involves an electron diffusive process. A
post selection procedure represented by the detection of charge
pairs in a region of the chain equidistant from Alice and Bob,
allows, if successful, the reconstruction on the Bob site of the
unknown state initially encoded by Alice. The peculiar feature of
the model is that the introduction of a trapped magnetic field
strongly improves the process efficiency.
\end{abstract}
\maketitle

\section{Introduction}

In the framework of quantum information \cite{nielsen} a key task is
represented by the ability of manipulating a quantum state in order
to realize a desired logical operation or to transfer it from one
location to another inside a quantum processor. An example of
channel proposed for quantum communication is represented by a spin
chain \cite{bose}. Here the information is encoded in a site by
rotating a spin state of an Heisenberg ferromagnet. The presence of
an unaligned spin creates a diffusion that allows to reconstruct the
information (i.e. the spin state) on a different site with a good
fidelity. Such a channel is also useful to transfer spin
entanglement, as showed in Ref. \cite{Subrahmanyam}. An $XY$
Hamiltonian permits a state transfer over long distances with
fidelity $1$ \cite{ekert}.

A quantum channel is defined as the evolution of a system from the
initial configuration to the final one.

The teleportation \cite{teleportation} is a particular kind of
quantum channel which exploits an entangled state shared by sender
(Alice) and receiver (Bob), plus local measurements and classical
communication.

Experimental implementations of teleportation protocol have been
reported in quantum optics \cite{optics1,optics2,optics3}, NMR
\cite{NMR} and very recently in atom physics \cite{blatt,knill}, and
proposals for solid state are currently object of interest \cite
{beenakker,feinberg,lidar,reina,0312112}.

This paper is devoted to the introduction of a model which describes
a teleportation on a quantum dot (QD) chain based on a diffusive
mechanisms. An unexpected feature of the model is that the success
probability of the protocol is enhanced through the use of a trapped
magnetic field proposed to improve the efficiency of the scheme.

The paper is articulated as follows. In section \ref{II} we
introduce the teleportation by extending the usual argument to the
effect of a time evolution of the system and introducing a mechanism
which implies a probabilistic behavior of the process. In section
\ref{III} is discussed an hamiltonian model which behaves as a
channel for diffusive teleportation. The presence of a trapped
magnetic field is shown to improve the efficiency of the process.
Last section will be devoted to conclusions.

\section{Teleportation under time evolution\label{II}}

Generally speaking, the teleportation works as follows. An unknown
state is encoded on a quantum two level system (qubit) in the
Alice's site: $\left| A\right\rangle =\alpha \left|
A_{1}\right\rangle +\beta \left| A_{2}\right\rangle $. Bob, being
far from Alice, wants to receive the qubit, preserving the
information contained thereinto, by exchanging with Alice only
classical information. To do it, they need to share an EPR state. To
complete the process, Alice performs a measurement operation (the so
called Bell measurement) on its own qubit and one component of the
entangled state. The result is then transmitted to Bob by way of
classical bits, and he can choose a proper unitary transformation to
apply on the second part of the shared EPR pair, recovering in such
a way the unknown state without need to know it.

Initially the system can be though as composed of an Alice's state
$\left| A\right\rangle $, a Bob's state and an intermediary system
state, entangled with the second one:
\begin{equation}
\left| \Psi \right\rangle =\left| A\right\rangle \otimes \left|
S,B\right\rangle =\left( \alpha \left| A_{1}\right\rangle +\beta
\left| A_{2}\right\rangle \right) \otimes \left( \left|
S_{1},B_{1}\right\rangle +\left| S_{2},B_{2}\right\rangle \right)
\label{1}
\end{equation}
where $|\alpha|^2+|\beta|^2=1$, and the Hamiltonian determining the
evolution is $H=H\left( A\otimes S\otimes B\right) $. Next, we
suppose that the time evolution involves just the subsystem $A
\otimes S$, while $B$ evolves in a trivial way. Labelling with
$\left| \Phi _{k}\right\rangle $ the energy eigenvectors of
$A\otimes S$
and with $E_{k}$\ the respective eigenvalues, fixing $\hbar =1$, we write $%
\left| A_{i},S_{j}\right\rangle
_{t}=\sum_{k}a_{k}^{ij}e^{-iE_{k}t}\left| \Phi _{k}\right\rangle $
(with $i,j=1,2$) and
\begin{equation}
\left| \Psi \right\rangle =\sum_{k}e^{-iE_{k}t}\left| \Phi
_{k}\right\rangle \left[ \left( \alpha a_{k}^{11}+\beta
a_{k}^{21}\right) \left| B_{1}\right\rangle +\left( \alpha
a_{k}^{12}+\beta a_{k}^{22}\right) \left| B_{2}\right\rangle \right]
\label{ij}
\end{equation}

The coefficients $a_{k}^{ij}$ are supposed to be known.

By an energy measurement performed on $A \otimes S$ Bob's state
collapses in a coherent superposition of $\left| B_{1}\right\rangle
$ and $\left| B_{2}\right\rangle $ with weights $\left( \alpha
a_{k}^{11}+\beta a_{k}^{21}\right) $ and $\left(
\alpha a_{k}^{12}+\beta a_{k}^{22}\right) $ connected to $\alpha $ and $%
\beta $ by a $k$-dependent transformation. In order to preserve the
universality of the protocol the transformation to implement to
recover the incoming Alice's state has to be independent from
$\alpha $ and $\beta $. In general cases this requirement is not
satisfied and the teleportation by energy measurement does not work.

The obstacle above illustrated is related to the difficulty of
realizing a full Bell measurement, which can be partially solved by
a partial Bell measurement\ \cite{pbm}. Alternatively, we propose
the idea to create a system which automatically excludes some state
as output result. As an example, we would write the overall state
introduced in Eq. \ref{1} as
\begin{equation}
\left| \Psi \right\rangle =\alpha \left| A_{1},S_{1}\right\rangle
\left| B_{1}\right\rangle +\beta \left| A_{2},S_{2}\right\rangle
\left| B_{2}\right\rangle  \label{reduced}
\end{equation}
Moreover, if the system is initially prepared in its ground state,
an energy selection can acts in the desired manner allowing to some
state to be more probably populated than some other.

When the simplification applies, Eq. \ref{ij} reduces to
\begin{equation}
\left| \Psi \right\rangle =\sum_{k}e^{-iE_{k}t}\left( \alpha
a_{k}^{11}\left| B_{1}\right\rangle +\beta a_{k}^{22}\left|
B_{2}\right\rangle \right) \left| \Phi _{k}\right\rangle
\label{alfa}
\end{equation}

From Eq. \ref{alfa} we note that the universality of the
teleportation can be observed if and only if $a_{1k}$ and $a_{2k}$
have the same modulus and differ only for a phase factor, i.e.$\
a_{1k}=e^{i\varphi _{k}}a_{2k}$ for each $k$. If the Hamiltonian
$H\left( AS\right) $ exhibits a non-degenerate spectrum Alice sends
to Bob the information about the measured energy level as classical
bit and finally he can do the conditional unitary operation to
completely reconstruct the incoming unknown state. Otherwise, if
$a_{1k}$ and $a_{2k}$ are connected by a more complicated relation,
there is no way to extract some useful information from an energy
measurement without knowing $\alpha $ and $\beta $ and the procedure
fails. If the condition $ a_{1k}=e^{i\varphi _{k}}a_{2k}$ is
satisfied just for some $k$ we deal with a teleportation protocol
characterized by a success probability less than $1$.

The situation above described is not the more general. Effectively,
to realize an energy measurement could be hard, and the difficulty
increases with the number of states implied in the evolution. The
measurement can be performed in a different basis, and a favorable
choice is often represented by the computational basis.

Writing a generic element $\left| \tilde{\Phi}_{l}\right\rangle $ of
the new set of orthogonal states as combination of energy
eigenstates, Eq. \ref{alfa} becomes
\begin{equation}
\left| \Psi \right\rangle =\sum_{l}\left( \alpha A_{1l}\left(
t\right) \left| \tilde{\Phi}_{l}\right\rangle \left|
B_{1}\right\rangle +\beta A_{2l}\left( t\right) \left|
\tilde{\Phi}_{l}\right\rangle \left| B_{2}\right\rangle \right)
\end{equation}
where $A_{il}\left( t\right) =\sum_{k}e^{-iE_{k}t}a_{ik}b_{kl}$ and $%
b_{kl}=\left\langle \tilde{\Phi}_{l}|\Phi _{k}\right\rangle $.

Now the condition to fulfil to deal with a deterministic teleportation is $%
A_{1l}\left( t\right) =e^{i\varphi _{l}}A_{2l}\left( t\right) $ and
the measurement time is then meaningful. Hence, if for certain times
this relation is satisfied, it will be sufficient to perform the \
Bell measurement at the right time.

However, this is not enough to ensure the protocol to be
deterministic. In the computational basis the information derived
from a measurement is expressed as ``yes'' or ``not'' and the space
dimension \ $d$ plays an important role. If only two are the
possible populated states, a measurement will be sufficient to
identify the right unitary transformation to perform over the Bob's
qubit, while this is not possible if $d>2$. A complete description
of a near deterministic, time dependent, teleportation scheme is
done elsewhere \cite{0312112}. Here we are interested to a model
which represents an example of probabilistic teleportation.

\section{The model\label{III}}

The model is represented by a chain of localized levels (QDs) with
double occupation allowed. Empty and doubly occupied dots are
energetically
degenerate, while single occupation is characterized by an energy $\epsilon $%
. Furthermore, we suppose that no electrons are initially present in
the chain. The qubit is defined as the coherent superposition of
vacuum and double occupation state in a single QD. Being possible
also single occupation, we deal with an ``open'' two level system.

The unknown state to teleport is encoded on a QD (representing
Alice) external to the chain as
\begin{equation}
\left| \psi \right\rangle =\alpha \left| 0\right\rangle +\beta
\left| \uparrow \downarrow \right\rangle
\end{equation}
where $\uparrow $ stays for spin up and $\downarrow $ stays for spin
down. A way to obtain this state is represented by the interaction
of the dot initially in its vacuum state with a superconductive
lead.

Bob is located in another external QD which is locally coupled with
an element of the chain.

The interaction between Bob and the chain dot is described by the
Hamiltonian
\begin{equation}
H=H_{0}+H_{I}  \label{acca}
\end{equation}
with
\begin{equation}
H_{0}=-2\epsilon \sum_{i=1}^{2}n_{i\uparrow }n_{i\downarrow
}+\epsilon \sum_{i=1}^{2}\left( n_{i\uparrow }+n_{i\downarrow
}\right)
\end{equation}
and
\begin{equation}
H_{I}=-w\sum_{\sigma }\left( c_{1,\sigma }^{\dagger }c_{2,\sigma
}+h.c.\right)
\end{equation}
where $c_{i,\sigma }$ $\left( c_{i,\sigma }^{\dagger }\right) $ is
the annihilation (creation) operator on dot $i$ (e.g. the dot $1$ is
Bob, the dot $2$ is the chain element) and $\sigma $ is the spin
index.

If an electron pair is prepared on Bob's dot, the incoming state is
$\left| \Phi _{1}\left( t=0\right) \right\rangle =\left| \uparrow
\downarrow ,0\right\rangle $. The other possible states are $\left|
\Phi _{2}\left(
t=0\right) \right\rangle =\left| 0,\uparrow \downarrow \right\rangle $, $%
\left| \Psi _{1}\left( t=0\right) \right\rangle =2^{-1/2}\left(
\left| \uparrow ,\downarrow \right\rangle +\left| \downarrow
,\uparrow \right\rangle \right) $, and $\left| \Psi _{2}\left(
t=0\right) \right\rangle =2^{-1/2}\left( \left| \uparrow ,\downarrow
\right\rangle -\left| \downarrow ,\uparrow \right\rangle \right) $.

In the limit $\epsilon >>w$ the state $\left| \Phi _{1}\left(
t=0\right) \right\rangle $ evolves, aside from a global phase
factor, in $\left| \Phi _{1}\left( t\right) \right\rangle =\left[
\cos \frac{\omega t}{2}\left| \Phi _{1}\left( t=0\right)
\right\rangle +i\sin \frac{\omega t}{2}\left| \Phi _{2}\left(
t=0\right) \right\rangle \right] +O(w/\epsilon )$, where $\omega
=\epsilon -\sqrt{\left( \epsilon ^{2}+16w^{2}\right) }$. If the
tunneling is switched off when $\omega t/2=\pi /4$ a maximally
entangled state is then generated.

After entanglement is created, the whole state describing Alice, the
chain and Bob is
\begin{equation}
\left| \psi \right\rangle =\frac{1}{2}\left( \alpha \left|
0\right\rangle _{A}\left| 0\right\rangle _{C}\left| \uparrow
\downarrow \right\rangle _{B}+\beta \left| \uparrow \downarrow
\right\rangle _{A}\left| 0\right\rangle _{C}\left| \uparrow
\downarrow \right\rangle _{B}+\alpha \left| 0\right\rangle
_{A}\left| \uparrow \downarrow \right\rangle _{C}\left|
0\right\rangle _{B}+\beta \left| \uparrow \downarrow \right\rangle
_{A}\left| \uparrow \downarrow \right\rangle _{C}\left|
0\right\rangle _{B}\right)  \label{9}
\end{equation}
where subscript $A$ denotes Alice'site, $B$ is Bob's site and $C$ is
the chain site nearest to Bob.

If Alice's dot is embedded in the chain and a sort of chemical
potential is added to the Hamiltonian in order to limit to one the
number of electron pairs on the enlarged chain, we obtain
\begin{equation}
\left| \psi \right\rangle =\alpha \left| 0\right\rangle _{A}\left|
\uparrow \downarrow \right\rangle _{C}\left| 0\right\rangle
_{B}+\beta \left| \uparrow \downarrow \right\rangle _{A}\left|
0\right\rangle _{C}\left| \uparrow \downarrow \right\rangle _{B}
\label{diffusive}
\end{equation}
in analogy with Eq. \ref{reduced}.

If the Hamiltonian introduced in Eq. \ref{acca} is now turned on for
all the elements in the chain, Bob is excluded, the states $\left|
0\right\rangle _{A}\left| \uparrow \downarrow \right\rangle _{C}$
and $\left| \uparrow \downarrow \right\rangle _{A}\left|
0\right\rangle _{C}$ \ experience a time evolution that consists in
a diffusion of the localized pair around the ring.

The diffusion is studied by the introduction of $\left| \Psi
_{l,m}\right\rangle $, describing the presence of an electron with
spin up
(down) on the site $l$($m$) and its Fourier transform $\left| \tilde{\Psi}%
_{k,q}\right\rangle $

\begin{equation}
\left| \tilde{\Psi}_{k,q}\right\rangle =\frac{1}{N}\sum_{l,m}\left|
\Psi _{l,m}\right\rangle e^{ikl}e^{imq}
\end{equation}
If we suppose that Alice is located in the site labeled with $0$ and
Bob is close to the $m$-th site, the initial state is
\begin{equation}
\left| \Psi \left( t=0\right) \right\rangle =\alpha \left| \Psi
_{m,m}\left( t=0\right) \right\rangle \left| 0\right\rangle
_{B}+\beta \left| \Psi _{0,0}\left( t=0\right) \right\rangle \left|
\uparrow \downarrow \right\rangle _{B}
\end{equation}
In the complex Laplace space the evolution is given by
\begin{gather}
\left( \omega -\epsilon \right) \left| \Psi _{l,m}\left( \omega
\right) \right\rangle =\left| \Psi _{l,m}\left( t=0\right)
\right\rangle -  \nonumber
\\
w\left[ \left| \Psi _{l+1,m}\left( \omega \right) \right\rangle
+\left| \Psi _{l,m-1}\left( \omega \right) \right\rangle +\left|
\Psi _{l-1,m}\left( \omega \right) \right\rangle +\left| \Psi
_{l,m+1}\left( \omega \right) \right\rangle \right] -\epsilon \left|
\Psi _{l,m}\left( \omega \right) \right\rangle \delta _{l,m}
\end{gather}
which is derived taking into account that states with double
occupation on a site have zero potential energy, and from which
follows
\begin{equation}
\left| \Psi _{m,m}\left( \omega \right) \right\rangle =\frac{1}{N}\sum_{k,q}%
\frac{e^{-i\left( k+q\right) m}\left| \tilde{\Psi}_{k,q}\left(
t=0\right) \right\rangle -\frac{\epsilon }{N}\sum_{l}\left| \Psi
_{l,l}\left( \omega \right) \right\rangle e^{i\left( l-m\right)
\left( k+q\right) }}{\omega -\epsilon +2w\left( \cos k+\cos q\right)
}  \label{mm}
\end{equation}
If a trapped magnetic field is introduced, because of the
Aharonov-Bohm effect \cite{aharonov}, the electron bands are shifted
by a quantity $\Phi =e\Phi^{\prime}/\hbar $, being $\Phi^{\prime}$
the flux of the magnetic field and $e$ the electron charge.

The Bell measurement is performed by selecting a site $l_{0}$ (from
symmetry reasons it will be natural to choose the intermediate dot
between $0\,$\ and $m$) and asking to find two charges here
localized: the Bob's state is then reduced to
\begin{equation}
\left| B\left( t\right) \right\rangle =\alpha f_{l_{0},m}\left(
t\right) \left| 0\right\rangle _{B}+\beta f_{l_{0},0}\left( t\right)
\left| \uparrow \downarrow \right\rangle _{B}  \label{bconf}
\end{equation}

where $f_{l_{0},p}\left( t\right) =\left\langle \Psi
_{l_{0},l_{0}}\left( t=0\right) |\Psi _{p,p}\left( t\right)
\right\rangle \,$($p=0,m$).

After some algebraic manipulation, whose details are given in
appendix A, we find
\begin{equation}
f_{l_{0},p}\left( \omega \right) =\frac{1}{\epsilon
N}\sum_{k}e^{+i\left( l_{0}-p\right) k}\frac{\Lambda \left( k,\omega
\right) }{1+\Lambda \left( k,\omega \right) }
\end{equation}
with
\begin{equation}
\Lambda \left( k,\omega \right) =\frac{\epsilon }{\sqrt{\left(
\omega -\epsilon \right) ^{2}-\left( 4w\cos \left( \frac{k}{2}-\Phi
\right) \right) ^{2}}}  \label{lambda}
\end{equation}
or
\[
1+\Lambda \left( k,\omega \right) =\frac{\epsilon }{\sqrt{\left(
\omega -\epsilon \right) ^{2}-\left( 4w\cos \left( \frac{k}{2}-\Phi
\right) \right) ^{2}}+\epsilon }
\]
Here we note that the cut deriving from the square root gives a
negligible
contribute and that a pole is present in $1+\Lambda \left( k,\omega \right) $%
\ only if $\epsilon <0$. Then, the two level of qubit encoding
represent excited states of the system.

Quite naturally, the projection can be performed on a region of
finite extension around $l_{0}$. If a gaussian form factor of width
$\sigma $ is used, we get
\begin{equation}
f_{l_{0},m}\left( t\right) =\frac{1}{N}\sum_{k}e^{ikl_{0}}e^{-\frac{%
k^{2}\sigma ^{2}}{2}}\frac{e^{i\epsilon t}}{2\pi i}\oint d\omega \frac{%
e^{i\omega t}\left[ \sqrt{\omega ^{2}-8w^{2}\left[ 1+\cos \left(
k-2\Phi \right) \right] }-\epsilon \right] }{\omega
^{2}-8w^{2}\left[ 1+\cos \left( k-2\Phi \right) \right] -\epsilon
^{2}}
\end{equation}
and
\begin{equation}
f_{l_{0},0}\left( t\right) =\frac{1}{N}\sum_{k}e^{-ikl_{0}}e^{-\frac{%
k^{2}\sigma ^{2}}{2}}\frac{e^{i\epsilon t}}{2\pi i}\int_{C}d\omega \frac{%
e^{i\omega t}\left[ \sqrt{\omega ^{2}-8w^{2}\left[ 1+\cos \left(
k-2\Phi \right) \right] }-\epsilon \right] }{\omega
^{2}-8w^{2}\left[ 1+\cos \left( k-2\Phi \right) \right] -\epsilon
^{2}}
\end{equation}
We note that the gaussian distribution makes relevant only small values of $%
k $. The roots of the denominator of the latter equations are
\begin{equation}
\omega _{\pm }=\pm \sqrt{\epsilon ^{2}+8w^{2}\left[ 1+\cos \left(
k-2\Phi \right) \right] }  \label{cosine}
\end{equation}
and, defining $\epsilon _{\pm }=\epsilon +\omega _{\pm }$, are
expressible as $\epsilon _{\pm }\simeq \epsilon _{\pm }\left(
k=0\right) +\epsilon _{\pm }^{\prime }\left( k=0\right) k+\epsilon
_{\pm }^{\prime \prime }\left( k=0\right) \frac{k^{2}}{2}$ with the
symmetry properties $\epsilon _{+}^{\prime \prime }\left( 0\right)
=-\epsilon _{-}^{\prime \prime }\left( 0\right) $ and $\epsilon
_{+}^{\prime }\left( 0\right) =-\epsilon _{-}^{\prime }\left(
0\right) $. If the trapped field were not present, the linear term
in $k$ should be zero, due to the cosine dependence.

By inverse Laplace transform we have
\begin{eqnarray}
f& _{l_{0},m}\left( t\right) =\frac{2}{N}\sum_{k}\frac{\left|
\epsilon \right| }{\epsilon _{+}-\epsilon _{-}}\{\exp \left[
i\epsilon _{+}\left(
0\right) t+ik\left[ \epsilon _{+}^{\prime }\left( 0\right) t-l_{0}\right] -%
\frac{k^{2}\lambda \left( t\right) }{2}\right] -  \nonumber \\
& -\exp \left[ i\epsilon _{-}\left( 0\right) t+ik\left[ -\epsilon
_{+}^{\prime }\left( 0\right) t-l_{0}\right] -\frac{k^{2}\lambda
^{\ast }\left( t\right) }{2}\right] \}  \label{fm}
\end{eqnarray}
and
\begin{eqnarray}
f_{l_{0},0}\left( t\right) =\frac{2}{N}\sum_{k}\frac{\left| \epsilon
\right| }{\epsilon _{+}-\epsilon _{-}} &\{&\exp \left[ i\epsilon
_{+}\left( 0\right)
t+ik\left[ \epsilon _{+}^{\prime }\left( 0\right) t+l_{0}\right] -\frac{%
k^{2}\lambda \left( t\right) }{2}\right] -  \nonumber \\
&&-\exp \left[ i\epsilon _{-}\left( 0\right) t+ik\left[ -\epsilon
_{+}^{\prime }\left( 0\right) t+l_{0}\right] -\frac{k^{2}\lambda
^{\ast }\left( t\right) }{2}\right] \}  \label{f0}
\end{eqnarray}
with $\lambda \left( t\right) =\sigma ^{2}+i\epsilon _{+}^{\prime
\prime }\left( 0\right) t$.

The linear term in $k$ is responsible of a quick attenuation because
of interference of different wavepacket components.

The introduction of the trapped magnetic field eliminates this
attenuation: without the field the term $\epsilon _{+}^{\prime
}\left( 0\right) $ is not present, and there is no way to bring to
zero the rapid diffusion term. On
the other hand, in the presence of the field, by choosing a proper time ($%
t_{M}=l_{0}/\epsilon _{+}^{\prime }\left( 0\right) $) for the
projective measurement, one of the linear terms in $k$ cancels both
in equation \ref{fm} and in equation \ref{f0}, giving rise to a
coherent propagation of the two electron wavepacket. The presence of
the term proportional to $k^{2}$ implies a broadening of the
gaussian wave function which can be optimized by a suitable choice
of the intensity of the magnetic field. Intuitively, the two
exponentials in each of latter equations represent
counterpropagating terms. The magnetic field has the effect to keep
the coherence of peaks. If usually two peaks starting from different
points are subject to a inessential global effect on the phase, the
peculiar structure above derived allows to superpose two peaks
starting form opposite point in the middle of the chain.

After the projection Bob's state is then

\begin{eqnarray}
\left| B\left( t_{M}\right) \right\rangle &=&\alpha \frac{2}{N}\sum_{k}\frac{%
\left| \epsilon \right| }{\epsilon _{+}-\epsilon _{-}}\exp \left(
i\epsilon _{+}\left( 0\right) t_{M}-\frac{k^{2}\lambda \left(
t_{M}\right) }{2}\right)
\left| 0\right\rangle _{B}-  \nonumber \\
&&-\beta \frac{2}{N}\sum_{k}\frac{\left| \epsilon \right| }{\epsilon
_{+}-\epsilon _{-}}\left[ \exp \left( i\epsilon _{-}\left( 0\right) t_{M}-%
\frac{k^{2}\lambda ^{\ast }\left( t_{M}\right) }{2}\right) \right]
\left| \uparrow \downarrow \right\rangle _{B}
\end{eqnarray}
Performing the sum over $k$ imposing $k=0$ in the difference
$(\epsilon _{+}-\epsilon _{-})$ we get
\begin{equation}
\left| B\left( t_{M}\right) \right\rangle =e^{i\left[ \left(
\epsilon -\epsilon _{0}\right) \text{\ }t_{M}+\theta \left(
t_{M}\right) \right]
}\left| f\left( t_{M}\right) \right| \left\{ \alpha e^{i\left[ 2\epsilon _{0}%
\text{\ }t_{M}-2\theta \left( t_{M}\right) +\frac{\pi }{2}\right]
}\left| 0\right\rangle _{B}+\beta \left| \uparrow \downarrow
\right\rangle _{B}\right\}  \label{btm}
\end{equation}
where
\begin{equation}
f\left( t_{M}\right) \simeq \frac{1}{\sqrt{2\pi }}\frac{1}{\lambda
^{\ast }\left( t_{M}\right) }\frac{\left| \epsilon \right|
}{\epsilon _{0}}
\end{equation}
with $f\left( t_{M}\right) =\left| f\left( t_{M}\right) \right|
e^{i\theta \left( t_{M}\right) }$ and $\epsilon _{0}=2\sqrt{\epsilon
^{2}+8w^{2}\left( 1+\cos 2\Phi \right) }$.

From Eq. \ref{btm} we learn that the state transferred to Bob is the
same initially encoded by Alice, apart from an inessential global
phase factor, a\ relative phase factor whose effects are eliminable
by a deterministic unitary operation, known {\em a priori},
different from the classically driven unitary rotation in the
standard teleportation, and a term ($\left| f\left( t_{M}\right)
\right| $) which creates attenuation: the scheme efficiency is then
$\left| f\left( t_{M}\right) \right| ^{2}$.

Writing explicitly
\begin{equation}
\left| f\left( t_{M}\right) \right| =\frac{1}{\sqrt{2\pi
}}\frac{\left|
\epsilon \right| }{\epsilon _{0}}\frac{1}{\sqrt{\sigma ^{4}+t_{M}^{2}\frac{%
16w^{4}}{\epsilon _{0}^{2}}\cos ^{2}2\Phi }}
\end{equation}
we note that the minimum attenuation is reached when $\Phi =\pi /4$
which implies\ \ $t_{M}=l_{0}\epsilon _{0}/4w^{2}$.

In the usual teleportation, after the Bell measurement, a classical
communication related to a unitary transformation is needed to
warrant against superluminar information transfer. Here, apparently,
the transfer is realized independently from the unitary operation.
Actually, due to the probabilistic nature of the process, Bob has to
know, through a classical channel, if Alice's measurement were
successful, and only in case of affirmative answer he is sure to be
in possess of the right unknown quantum state.

If the constraint of only one electron pair propagating on the chain
is relaxed, the probability amplitude of observing a pair in the
middle point is modified by the contribution of configurations where
two other spins are somewhere in the chain (the last term in Eq.
\ref{9}). A peculiar feature of the model is that this contribution
is not enhanced by the magnetic field and its effects on the scheme
is negligible. In appendix B we shall show analytically this
argument.

\section{Conclusions}

We have studied the teleportation protocol introducing the
possibility that some subcomponent is subject to time evolution. A
model which fulfils this characteristic has been introduced by means
of a chain of interacting quantum dots. To improve the protocol
efficiency, a trapped magnetic field has been introduced. We
observed that a coherent, although attenuated, propagation from
Alice and Bob of two counterpropagating wavepackets, plus a
postselection measurement which determines the presence of a
particle pair in the middle point at a selected time permits to
realize the teleportation. The model proposed is an ''open quantum
system'' with the qubit defined on two excited states. A natural
limit is represented by unavoidable attenuation due to diffusion and
fragility with respect to decoherence eventually due to interaction
with an external phonon bath, since the mechanism works exploiting
local excited states on each QD.

\renewcommand{\theequation}{A-\arabic{equation}}
\setcounter{equation}{0} 

\section*{APPENDIX A: DERIVATION OF COEFFICIENTS IN EQ. \ref{bconf}\label{A}}



We have to calculate the coefficient
\begin{equation}
f_{l,m}\left( \omega \right) =\left\langle \Psi _{l,l}\left(
t=0\right) |\Psi _{m,m}\left( \omega \right) \right\rangle \text{\ \
\ }
\end{equation}
where $l,m$ are now two generic sites, starting from Eq. \ref{mm},
which we recall for convenience:
\begin{equation}
\left| \Psi _{m,m}\left( \omega \right) \right\rangle =\frac{1}{N}\sum_{k,q}%
\frac{e^{-i\left( k+q\right) m}\left| \tilde{\Psi}_{k,q}\left(
t=0\right) \right\rangle -\frac{\epsilon }{N}\sum_{l}\left| \Psi
_{l,l}\left( \omega \right) \right\rangle e^{i\left( l-m\right)
\left( k+q\right) }}{\omega -\epsilon +2w\left( \cos k+\cos q\right)
}  \label{a1}
\end{equation}
Noting that the inner product $\left\langle \Psi _{l,l}\left( t=0\right) |%
\tilde{\Psi}_{k,q}\left( t=0\right) \right\rangle $ is equal to
$e^{+i\left( k+q\right) l}/N$, Eq. \ref{a1} becomes
\begin{equation}
f_{l,m}\left( \omega \right) =\frac{1}{N}\sum_{k,q}\frac{1}{\omega
-\epsilon +2w\left( \cos k+\cos q\right) }\left[
\frac{1}{N}e^{-i\left( m-l\right) \left( k+q\right) }-\frac{\epsilon
}{N}\sum_{l^{\prime }}f_{l,l^{\prime
}}\left( \omega \right) e^{i\left( l^{\prime }-m\right) \left( k+q\right) }%
\right]
\end{equation}
or\qquad
\begin{equation}
f_{l,m}\left( \omega \right) =\frac{1}{N}\sum_{k,q}\frac{1}{\omega
-\epsilon
+2w\left[ \cos \left( k-q\right) +\cos q \right] }%
\left[ \frac{1}{N}e^{-i\left( m-l\right) k}-\frac{\epsilon }{N}%
\sum_{l^{\prime }}f_{l,l^{\prime }}\left( \omega \right) e^{i\left(
l^{\prime }-m\right) k}\right]
\end{equation}
By defining the quantity
\begin{equation}
f_{l}\left( k,\omega \right) =\frac{1}{\sqrt{N}}\sum_{l^{\prime
}}f_{l,l^{\prime }}\left( \omega \right) e^{ikl^{\prime }}
\end{equation}
we obtain the following identity:
\begin{equation}
f_{l}\left( k,\omega \right)
=\frac{1}{\sqrt{N}}\sum_{q}\frac{1}{\omega
-\epsilon +2w\left[ \cos \left( k-q\right) +\cos q\right] }\left[ \frac{1}{N}%
e^{+ilk}-\frac{\epsilon }{\sqrt{N}}f_{l}\left( k,\omega \right)
\right]
\end{equation}
which can be rewritten as
\begin{equation}
f_{l}\left( k,\omega \right) =\frac{1}{\epsilon \sqrt{N}}e^{+ilk}\frac{%
\Lambda \left( k,\omega \right) }{1+\Lambda \left( k,\omega \right)
}
\end{equation}
where
\begin{equation}
\Lambda \left( k,\omega \right) =\frac{\epsilon
}{N}\sum_{q}\frac{1}{\omega -\epsilon +4w\cos \left(
q-\frac{k}{2}\right) \cos \frac{k}{2}}
\end{equation}
After substitution, we get
\begin{equation}
f_{l,m}\left( \omega \right) =\frac{1}{\epsilon
N}\sum_{k}e^{+i\left( l-m\right) k}\frac{\Lambda \left( k,\omega
\right) }{1+\Lambda \left( k,\omega \right) }
\end{equation}
Equation \ref{lambda} is derived performing the sum over $q$ in the
continuous limit.

\renewcommand{\theequation}{B-\arabic{equation}} \setcounter{equation}{0}

\section*{APPENDIX B: TIME EVOLUTION OF TWO ELECTRON PAIRS}

In this appendix we discuss the case of time evolution of last term
in Eq. \ref{9}. The diffusion arises from an initial state where the
first electron pair is localized in the site $l$ and the second is
localized in the site $m$. The system state is described through its
Laplace transform $\left| \Psi _{l,m;n,p}\left( \omega \right)
\right\rangle $ which is subject to
\begin{gather}
\left( \omega -2\epsilon \right) \left| \Psi _{l,m;n,p}\left( \omega
\right) \right\rangle =\left| \Psi _{l,m;n,p}\left( t=0\right)
\right\rangle -2\epsilon \left| \Psi _{l,m;n,p}\left( \omega \right)
\right\rangle \left(
\delta _{l,n}\delta _{m,p}+\delta _{l,p}\delta _{m,n}\right) -  \nonumber \\
-\epsilon \left| \Psi _{l,m;n,p}\left( \omega \right) \right\rangle
\left( \delta _{l,n}+\delta _{l,p}+\delta _{m,n}+\delta
_{m,p}\right) -  \nonumber
\\
-w[\left| \Psi _{l-1,m;n,p}\left( \omega \right) \right\rangle
+\left| \Psi _{l+1,m;n,p}\left( \omega \right) \right\rangle +\left|
\Psi _{l,m-1;n,p}\left( \omega \right) \right\rangle +\left| \Psi
_{l,m+1;n,p}\left( \omega \right) \right\rangle +  \nonumber \\
+\left| \Psi _{l,m;n-1,p}\left( \omega \right) \right\rangle +\left|
\Psi _{l,m;n+1,p}\left( \omega \right) \right\rangle +\left| \Psi
_{l,m;n,p-1}\left( \omega \right) \right\rangle +\left| \Psi
_{l,m;n,p+1}\left( \omega \right) \right\rangle ]
\end{gather}
obtained taking into account that states with different number of
paired electrons give a different contribution to the potential
energy.

By the introduction of Fourier transform, defined as
\begin{equation}
\left| \Psi _{k,q;r,s}\left( \omega \right) \right\rangle =\frac{1}{N^{2}}%
\sum_{l,m,n,p}e^{ikl}e^{iqm}e^{irn}e^{isp}\left| \Psi
_{l,m;n,p}\left( \omega \right) \right\rangle
\end{equation}
we are able to write
\begin{gather}
\left| \tilde{\Psi}_{k,q;r,s}\left( \omega \right) \right\rangle =\frac{1}{%
\omega -\epsilon \left( k,q,r,s\right) }\ast  \nonumber \\
\lbrack \left| \tilde{\Psi}_{k,q;r,s}\left( t=0\right) \right\rangle -\frac{%
\epsilon }{N^{2}}\sum_{l,m,p}\left| \Psi _{l,m;l,p}\left( \omega
\right)
\right\rangle e^{i\left( k+r\right) l}e^{iqm}e^{isp}-\frac{\epsilon }{N^{2}}%
\sum_{l,m,n}\left| \Psi _{l,m;n,l}\left( \omega \right)
\right\rangle
e^{i\left( k+s\right) l}e^{iqm}e^{irn}-  \nonumber \\
\frac{\epsilon }{N^{2}}\sum_{l,m,p}\left| \Psi _{l,m;m,p}\left(
\omega \right) \right\rangle e^{ikl}e^{i\left( q+r\right)
m}e^{isp}-\frac{\epsilon }{N^{2}}\sum_{l,m,n}\left| \Psi
_{l,m;n,m}\left( \omega \right)
\right\rangle e^{ikl}e^{i\left( qb+s\right) m}e^{irn}-  \nonumber \\
\frac{2\epsilon }{N^{2}}\sum_{l,m}\left| \Psi _{l,m;l,m}\left(
\omega
\right) \right\rangle e^{i\left( k+r\right) l}e^{i\left( q+s\right) m}-\frac{%
2\epsilon }{N^{2}}\sum_{l,m}\left| \Psi _{l,m;m,l}\left( \omega
\right) \right\rangle e^{i\left( k+s\right) l}e^{i\left( q+r\right)
m}]  \label{b2}
\end{gather}
having introduced $\epsilon \left( k,q,r,s\right) =\left[ 2\epsilon
-2w\left( \cos k+\cos q+\cos r+\cos s\right) \right] $.

Then, the projection on the site $l_{0}$\ of two pairs starting from
$l$ and $m$, which we call $f\left( l,m,l_{0}\right) $, is related
to the inner product $\left\langle \Psi _{l_{0},m^{\prime
};l_{0},p^{\prime }}|\Psi
_{l,m;l,m}\left( \omega \right) \right\rangle $ (being $m^{\prime }$\ and $%
p^{\prime }$\ the sites occupied by the other two electrons).
\begin{equation}
f\left( l,m,l_{0};\omega \right) =\sum_{m^{\prime },p^{\prime
}}\left\langle \Psi _{l_{0},m^{\prime };l_{0},p^{\prime }}|\Psi
_{l,m;l,m}\left( \omega \right) \right\rangle
\end{equation}
By applying Fourier transform we obtain
\begin{equation}
f\left( l,m,l_{0};\omega \right) =\frac{1}{N^{4}}\sum_{m^{\prime
},p^{\prime }}\sum_{k^{\prime },q^{\prime },r^{\prime },s^{\prime
}}\sum_{k,q,r,s}e^{-il_{0}\left( k^{\prime }+r^{\prime }\right)
}e^{-i\left( q^{\prime }m^{\prime }+s^{\prime }p^{\prime }\right)
}e^{il\left( k+r\right) }e^{im\left( q+s\right) }\left\langle
\tilde{\Psi}_{k^{\prime },q^{\prime };r^{\prime },s^{\prime
}}|\tilde{\Psi}_{k,q;r,s}\left( \omega \right) \right\rangle
\end{equation}
that, performing the sums over $m^{\prime }$ and $p^{\prime }$,
reduces to
\begin{equation}
f\left( l,m,l_{0};\omega \right) =\frac{1}{N^{2}}\sum_{k^{\prime
},r^{\prime },k,q,r,s}\left[ e^{-il_{0}\left( k^{\prime }+r^{\prime
}\right)
}e^{il\left( k+r\right) }e^{i\left( mq+ps\right) }\left\langle \tilde{\Psi}%
_{k^{\prime },0;r^{\prime },0}|\tilde{\Psi}_{k,q;r,s}\left( \omega
\right) \right\rangle \right]
\end{equation}
Defining
\begin{equation}
\left| \tilde{\Psi}_{1}\left( k,q,s;\omega \right) \right\rangle =\frac{1}{%
\sqrt{N}}\sum_{r}\left| \tilde{\Psi}_{k-r,q;r,s}\left( \omega
\right) \right\rangle
\end{equation}
\begin{equation}
\left| \tilde{\Psi}_{2}\left( k,q,r;\omega \right) \right\rangle =\frac{1}{%
\sqrt{N}}\sum_{s}\left| \tilde{\Psi}_{k-s,q;r,s}\left( \omega
\right) \right\rangle
\end{equation}
\begin{equation}
\left| \tilde{\Psi}_{3}\left( k,q,s;\omega \right) \right\rangle =\frac{1}{%
\sqrt{N}}\sum_{r}\left| \tilde{\Psi}_{k,q-r;r,s}\left( \omega
\right) \right\rangle
\end{equation}
\begin{equation}
\left| \tilde{\Psi}_{4}\left( k,q,r;\omega \right) \right\rangle =\frac{1}{%
\sqrt{N}}\sum_{s}\left| \tilde{\Psi}_{k,q-s;r,s}\left( \omega
\right) \right\rangle
\end{equation}
\begin{equation}
\left| \tilde{\Psi}_{5}\left( k,q;\omega \right) \right\rangle =\frac{1}{N}%
\sum_{r,s}\left| \tilde{\Psi}_{k-s,q-r,r,s}\left( \omega \right)
\right\rangle
\end{equation}
\begin{equation}
\left| \tilde{\Psi}_{6}\left( k,q;\omega \right) \right\rangle =\frac{1}{N}%
\sum_{r,s}\left| \tilde{\Psi}_{k-r,q-s,r,s}\left( \omega \right)
\right\rangle
\end{equation}
\bigskip we find that
\begin{gather}
f\left( l,m,l_{0};\omega \right) =\frac{1}{N}\sum_{k^{\prime },k,q}\frac{1}{%
1+2\Pi \left( k,q\right) }e^{-il_{0}k^{\prime }}e^{ilk}e^{imq}\frac{1}{N}%
\sum_{r,s}\frac{1}{\omega -\epsilon \left( k-s,q-r,r,s\right)
}\left\langle \tilde{\Psi}_{k^{\prime },0;r^{\prime },0}\left(
t=0\right) \right|
\nonumber \\
\{\left| \tilde{\Psi}_{k-s,q-r;r,s}\left( t=0\right) \right\rangle -\frac{%
\epsilon }{\sqrt{N}}\left| \tilde{\Psi}_{1}\left( k+r-s,q-r,s;\omega
\right) \right\rangle -\frac{\epsilon }{\sqrt{N}}\left|
\tilde{\Psi}_{2}\left(
k,q-r,r;\omega \right) \right\rangle -  \nonumber \\
\frac{\epsilon }{\sqrt{N}}\left| \tilde{\Psi}_{3}\left(
k-s,q,s;\omega
\right) \right\rangle -\frac{\epsilon }{\sqrt{N}}\left| \tilde{\Psi}%
_{4}\left( k-s,q-r+s,r;\omega \right) \right\rangle -\frac{2\epsilon }{N}%
\left| \tilde{\Psi}_{6}\left( k-s+r,q-r+s;\omega \right)
\right\rangle \} \label{flm}
\end{gather}

where
\begin{equation}
\Pi \left( k,q\right) =\frac{1}{N^{2}}\sum_{r,s}\frac{1}{\omega
-\epsilon \left( k-s,q-r,r,s\right) }
\end{equation}
Eq. \ref{flm} has now to be integrated in the complex plane. We note
that
the unique contribution giving rise to a pure pole is the inner product $%
\left\langle \tilde{\Psi}_{k^{\prime },0;r^{\prime },0}\left( t=0\right) |%
\tilde{\Psi}_{k-s,q-r;r,s}\left( t=0\right) \right\rangle $, because
all other terms present an integration over the pole and can be, in
a perturbative approach, neglected.

Hence, applying conservation rules due to index matching,
\begin{equation}
f\left( l,m,l_{0};\omega \right) =\frac{1}{N^{2}}\sum_{k^{\prime },k,q}\frac{%
e^{i\left( l-l_{0}\right) k}e^{i\left( m-l_{0}\right) q}}{1+2\Pi
\left( k,q\right) }\frac{1}{\omega -\epsilon \left( k,0,q,0\right) }
\end{equation}
Limiting ourselves to small values of $q$ and $k$, the pole is in
\begin{equation}
\omega _{0}=2\epsilon +2w\left( k^{2}+q^{2}\right)
\end{equation}
Then
\begin{equation}
f\left( l,m,l_{0};t\right) =\frac{1}{N^{2}}\sum_{k,q}\frac{%
e^{idk}e^{-idq}e^{i\omega _{0}t}}{\left[ 1+2\Pi \left( k,q,\omega
_{0}\right) \right] }
\end{equation}
where $d=l-l_{0}=l_{0}-m$.

The effect of the trapped magnetic field manifests itself giving for
the pole $\omega _{0}=2\epsilon +2w\left( 2\Phi ^{2}+2\Phi
k+k^{2}+2\Phi q+q^{2}\right) $.

Then the linear part in $q$ and $k$ is $\exp i\left[ d\left(
k-q\right) +4w\Phi \left( k+q\right) t\right] $. A proper choice of
time for projecting the state will permit only to eliminate one of
two terms responsible for
rapid diffusion, and the probability of finding two electrons on the site $%
l_{0}$ will be negligible.

\end{document}